\documentclass[conference]{IEEEtran}

\usepackage[utf8]{inputenc}
\usepackage[T1]{fontenc}
\usepackage{microtype}

\usepackage[keeplastbox,ancient]{flushend}

\usepackage{booktabs}
\usepackage{graphicx}
\usepackage{url}

\usepackage{amsmath}

\usepackage{pifont}
\newcommand{\cmark}{\ding{51}}%
\newcommand{\xmark}{\ding{55}}%

\usepackage{enumitem}

\usepackage{enumitem} 

\makeatletter
\def\ps@IEEEtitlepagestyle{%
  \def\@oddfoot{\mycopyrightnotice}%
  \def\@evenfoot{}%
}
\def\mycopyrightnotice{%
  {\begin{minipage}{\textwidth}
  \footnotesize \copyright 2020 IEEE. Personal use of this material is permitted. Permission from IEEE must be obtained for all other uses, in any current or future media, including reprinting\slash republishing this material for advertising or promotional purposes, creating new collective works, for resale or redistribution to servers or lists, or reuse of any copyrighted component of this work in other works.
  \end{minipage}
  }
  \gdef\mycopyrightnotice{}
}

\begin{document}

\title{Data-driven Risk Management for Requirements Engineering: An~Automated Approach based on Bayesian Networks}

\author{
\IEEEauthorblockN{Florian Wiesweg}
\IEEEauthorblockA{Technische Universit\"at Berlin\\
Berlin, Germany\\
florian.wiesweg@alumni.tu-berlin.de}
\and
\IEEEauthorblockN{Andreas Vogelsang}
\IEEEauthorblockA{Technische Universit\"at Berlin\\
Berlin, Germany\\
andreas.vogelsang@tu-berlin.de}
\and
\IEEEauthorblockN{Daniel Mendez}
\IEEEauthorblockA{Blekinge Institute of Technology and fortiss GmbH\\
Blekinge, Sweden and Munich, Germany\\
daniel.mendez@bth.se}
}

\maketitle

\begin{abstract}
Requirements Engineering (RE) is a means to reduce the risk of delivering a product that does not fulfill the stakeholders' needs. Therefore, a major challenge in RE is to decide how much RE is needed and what RE methods to apply. The quality of such decisions is strongly based on the RE expert's experience and expertise in carefully analyzing the context and current state of a project. Recent work, however, shows that lack of experience and qualification are common causes for problems in RE. We trained a series of Bayesian Networks on data from the NaPiRE survey to model relationships between RE problems, their causes, and effects in projects with different contextual characteristics. These models were used to conduct (1) a post-mortem (diagnostic) analysis, deriving probable causes of sub-optimal RE performance, and (2) to conduct a preventive analysis, predicting probable issues a young project might encounter. The method was subject to a rigorous cross-validation procedure for both use cases before assessing its applicability to real-world scenarios with a case study.
\end{abstract}

\section{Introduction}

The purpose of Requirements Engineering (RE) is to elicit, document, analyze, and manage requirements to minimize the risk of delivering a system that does not meet the stakeholders' desires and needs~\cite{Glinz14}. 
Over the last 30 years, a number of methods, processes, tools, and best practices have been proposed to support this goal. 
However, there is no silver-bullet method or process that fits every project. 
In fact, a large part of the job of a requirements engineer in practice is to observe and analyze the context and current state of a project carefully and decide how much and what kind of RE is beneficial. 
As already addressed in the above-mentioned definition of RE, this decision is often a matter of controlling risks. 
Conducting RE tasks always comes with costs that ideally pay off in the sense that they lower a particular risk for a project~\cite{Femmer19}.
Making such decisions demands social and technical skills but also a lot of experience.
Recent studies have shown that lack of experience and lack of qualification of RE team members are the second and third most common causes for problems in RE (lack of time being the top cause)~\cite{mendezfernandezNamingPainRequirements2017}.
As a result, a number of projects fail either because of too little RE leading to stakeholder dissatisfaction or too much RE leading to high costs and developer frustration.

In this paper, we propose a data-driven approach to risk management in RE. 
Our goal is to predict RE problems, their causes, and effects for a given project. 
Intuitively, these three phenomena depend on each other in the sense that causes lead to problems, which result in certain effects. 
Bayesian Networks can be used to characterize such dependencies quantitatively by conditional probabilities and update the probability of certain phenomena when other phenomena are observed.
Therefore, we evaluated different versions of Bayesian Networks that model the relations between causes, problems, and effects in RE. 
We trained the models on data that was collected through two surveys with answers from 228 and 488 practitioners, respectively, about problems, causes, and effects encountered in real projects. These surveys also provide data on the context of the projects.

We use the trained models for the following two use cases:

\begin{itemize}
  \item {\bfseries Post-Mortem Analysis:} Given a set of problems and effects observed in a failing or failed project, the approach diagnoses the most likely causes leading to these issues (known as \textit{diagnostic reasoning} in literature~\cite{delaguilaBayesianNetworksEnhancement2016}).
  \item {\bfseries Preventive Analysis:} Given a set of causes and effects observed in a new or running project, the approach predicts the most likely problems to be faced (known as \textit{predictive reasoning} in literature~\cite{delaguilaBayesianNetworksEnhancement2016}).
\end{itemize}

We implemented the approach as an easily consumable web service, on which we based a graphical user interface to enter evidence and analyze the resulting predictions.

We performed two types of evaluations for our approach. Firstly, we performed cross-validation to compare the predictive power of different models. We achieved the best results for both use cases with surprisingly simple models, which ignore the causal structure implied by the original survey but include a set of context factors. For varying probability thresholds \(t \in \{0.3, 0.5, 0.7\}\), the best diagnostic model achieves recalls of 0.6, 0.48, 0.44 and precisions of 0.76, 0.92, 0.99, respectively. The best predictive model achieves recalls of 0.84, 0.69, 0.59 and precisions of 0.71, 0.89, 0.99.
A ranking-based output of the top-5 predictions results in a recall of 0.81 and a precision of 0.38 for the best diagnostic model and a recall of 0.73 and a precision of 0.71 for the best predictive model.

Secondly, we conducted a case study in industry to evaluate the external validity of the approach. We compared and discussed the predictions of the tool with the expectations of an RE expert for the diagnostic reasoning use case. Furthermore, we elicited feedback regarding the importance of recall vs. precision for the problem and how the tool should be tailored in detail to support practitioners best. In a nutshell, the case study showed that the method achieves good congruence between its predictions and the results expected by the expert, but requires additional tuning towards high precision.

We conclude that such data-driven approaches are very likely to be practical and advantageous, but that the remaining potentials in the underlying data and the user interface should be realized first.

\section{Related Work}

\subsection{NaPiRE Initiative} The survey data used for our analysis originates from the NaPiRE project, which was presented on several occasions~\cite{mendezfernandezNamingPainRequirements2017,mendezfernandezSupportingRequirementsEngineeringResearch2018,wagnerStatusQuoRequirements2019}. Formerly a German initiative, it has incorporated a variety of teams of other nationalities since its inception and is now supported by RE researchers from all over the world. Most analyses run on the data have so far been of a descriptive nature, e.g., comparing summary statistics from different countries~\cite{mendezfernandezNamingPainRequirements2015}, or trying to find the most prevalent problems, their causes, and their effects in RE projects~\cite{mendezfernandezNamingPainRequirements2017}. Two studies, however, applied Bayesian Networks to analyze the relationship between these items, with the purpose of either supporting the so-called \textit{Defect Causal Analysis}~\cite{kalinowskiSupportingDefectCausal2017} (the diagnostic reasoning use case) or allowing data-driven risk-management (the predictive reasoning use case)~\cite{mendezfernandezEvidenceBasedRiskManagement2018}. Both studies relied on the commercial Netica tool and lack a sophisticated validation procedure including an evaluation of their predictive power.

\subsection{Bayesian Networks in SE and RE} While certainly not a common tool for software engineers, Bayesian Networks have seen a variety of applications in Software and Requirements Engineering according to a mapping study~\cite{misirliBayesianNetworksEvidenceBased2014}, ranging (in declining importance) from software fault detection over software project management to design and testing. We would position this work in the second category, software project management. The survey also examines the methodological approach taken in the field: 80\% of the 117 works rely solely on categorical variables, just as we do, while empirical data is used in only about 45\% of the cases to learn the parameters of the model. In 24\% of the cases, the network layout is inferred from the data as well, which we do heuristically. According to a follow-up survey by the same authors~\cite{tosunSystematicLiteratureReview2017}, there has been a trend towards data-driven methods and continuous variables. These claims should, however, be seen in the light of the very low sample size of only 10 papers.

A survey of 20 studies matching our application in RE a bit more closely can be found in~\cite{delaguilaBayesianNetworksEnhancement2016}, confirming similar trends: the network layout is usually constructed manually, while some approaches use quantitative, data-driven methods for parameter learning. Two of these studies treat the more specific topic of the RE process: Tang~et~al.~\cite{tangRequirementEngineeringTechniques2009} use a Bayesian Network based on survey data to propose a set of requirements engineering techniques for different phases of the RE process. Nagy~et~al.~\cite{nagyBayesianBasedMethod2010} employ a network with a manually specified layout and parameters for development release planning and project health monitoring. None of the two approaches was systematically validated and different model designs were not evaluated.

\subsection{Use Cases} The diagnostic reasoning use case is inspired by concepts such as \textit{Root Cause Analysis} or \textit{Defect Causal Analysis} with original works stemming from the early 90s~\cite{cardDefectcausalAnalysisDrives1993,wilsonRootCauseAnalysis1993}, focusing mainly on quality management techniques. An interesting list of data-driven approaches is presented by Sol{\'e}~et~al.~\cite{soleSurveyModelsTechniques2017} (unfortunately only as a pre-print), including a variety of applications of Bayesian Networks to the topic.

The preventive analysis use case aids risk management by estimating likely problems. These risks can then be tackled by choosing matching RE methods, so it is possible to implement Just-in time-RE~\cite{Ernst12}, Just enough RE~\cite{Davis05}, or Value-oriented RE~\cite{Glinz08} in an empirically founded way. The model-based risk management tools available so far~\cite{asnarGoaldrivenRiskAssessment2011,grunskeQuantitativeRiskbasedSecurity2008} neither apply specifically to RE nor do they incorporate larger data sets.

\begin{figure}
  \centering
  \includegraphics[width=0.8\linewidth]{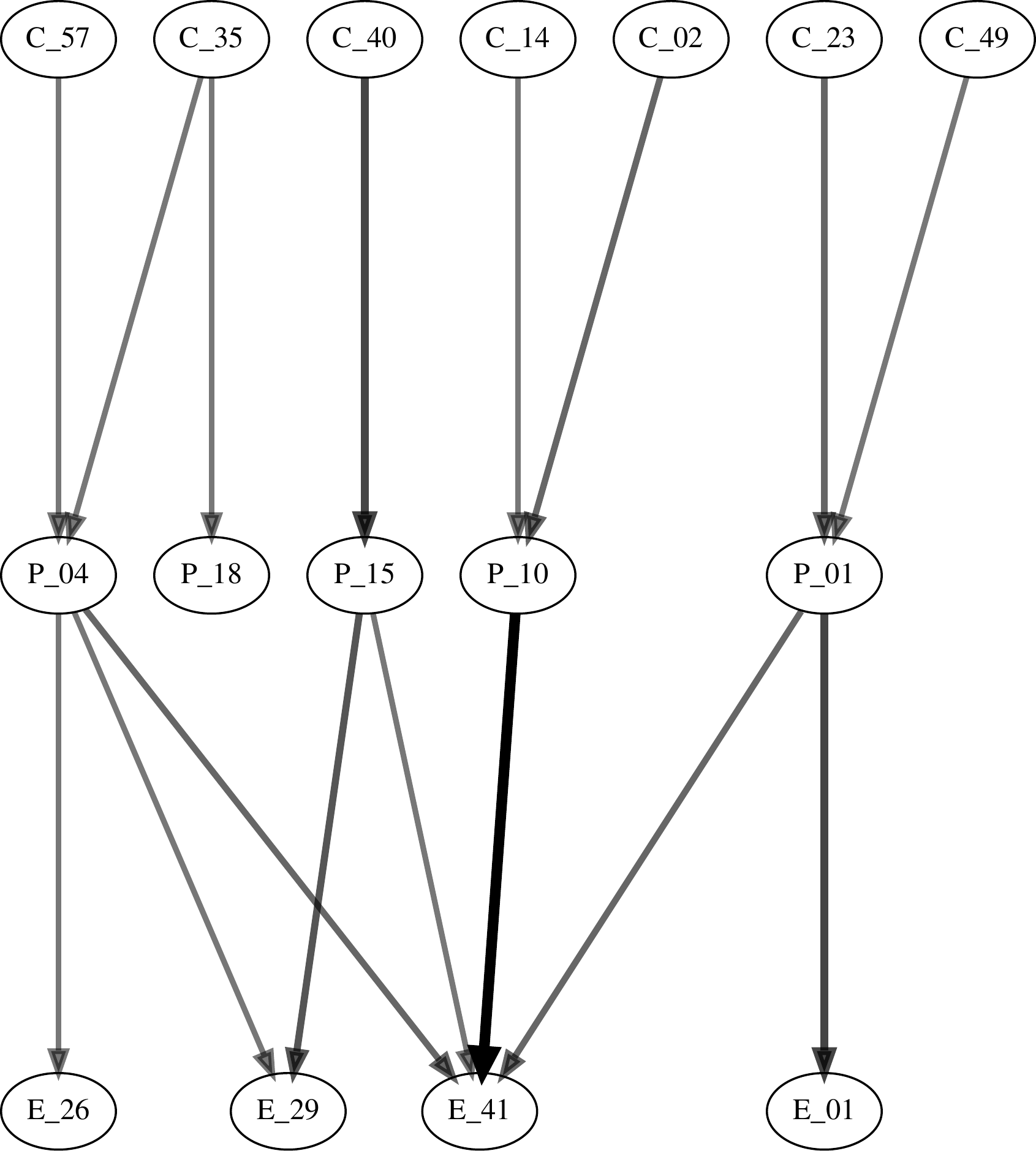}
  \caption[Visualization of a simple Bayesian Network]{Visualization of a Bayesian Network. Line thickness indicates how often the relation was mentioned by participants.\label{fig:example-graph}}
\end{figure}

\section{Bayesian Networks}

Bayesian Networks (also called belief networks) are part of a class of stochastic models called \textit{graphical models}, also including hidden Markov models and conditional random fields, which are popular in machine vision applications. They encode the joint stochastic distribution of a set of discrete and dependent random variables \({X = \{X_1, \ldots, X_n\}}\) in a directed acyclic graph such as the one depicted in Figure~\ref{fig:example-graph}. Each node in the graph represents a random variable \(X_i\) whose distribution is dependent on its predecessors \(pa(X_i)\), and this conditional distribution is stored along with the node. The explicit specification of dependencies strongly reduces the size of the parameter vector \(\boldsymbol{\theta}\): a toy network of \(n = 4\) binomial variables \(X_1, X_2, X_3, X_4\) with edges pointing from \(X_i\) to \(X_{i+1}\) would require a vector of \(2^n = 16\) scalars in a naive joint distribution specification, while a Bayesian Network formulation only requires \(2 \cdot n = 8\) scalars.

It follows from the above definition that a variable \(X_i\) in such a network is (conditionally) independent from any other variable in the graph given its predecessors \(pa(X_i)\), so the joint probability function of the model is

\begin{align*}
 P(X_1, \ldots, X_n) = \prod_i^n P(X_i\;|\;pa(X_i), \boldsymbol{\theta})
\end{align*}

Given \(m\) instantiations of all variables \(x_1^j, \ldots, x_n^j, 1 \leq j \leq m\) , it is thus possible to calculate the likelihood of such a model:

\begin{align*}
 L(\boldsymbol{\theta}\;|\;x^1, \ldots, x^m) = \prod_{j=1}^m P(X_1 = x_1^j, \ldots, X_n = x_n^j\;|\;\boldsymbol{\theta})
\end{align*}

The optimization of this function with (usually infeasible) analytical or (usually gradient-based) numerical methods yields the maximum likelihood point estimate of \(\boldsymbol{\theta}\)\footnote{The stochastic formulation of the problem enables the inference of confidence intervals for all parameters and the predictions, which might be an interesting extension of this work if practitioners show interest.}.

If the actual instantiations of any other random variables are known, this evidence \(E,\;{E \cap V_o = \emptyset}\) can be introduced into the network by manually modifying the conditional distributions at the respective nodes to generate a prediction tailored to the situation. \(v(X)\) being the set of all possible assignments to a set of random variables, the conditional probability is calculated as follows:

\begin{align*}
 P(V_o = v_o\;|\;E = e) &= \frac{P(V_o = v_o \wedge E = e)}{P(E = e)} \\
                      &= \frac{\sum_{v(X_i/E/V_o)} P(E = e, V_o = v_o)}{\sum_{v(X_i/E)} P(E = e, X_i = x_i)}
\end{align*}

This marginalization operation is more expensive than it would be for a naive joint distribution formulation, but the reduced memory requirements and statistical advantages of the smaller parameter space generally outweigh this concern. In addition, a variety of optimized approximate algorithms (such as Belief Propagation, or Gibbs Sampling, which we used) exploit the specific structure of the inference problem to reduce the overall computation time considerably.

For a more in-depth introduction to the theoretical foundations of Bayesian Networks, we suggest the textbook by Koski and Noble~\cite{koskiBayesianNetworksIntroduction2009}.

\section{Approach}
\subsection{Assumptions and Design Decisions}
Our contribution to RE risk management is an approach that is able to infer and report probabilities of certain RE problems, their effects, and their likely causes. The approach \emph{learns} this inference from a large set of observations recorded in other RE projects in the past. Conceptually, we assume that these three classes of phenomena are dependent in the sense that causes lead to problems, which result in effects. Therefore, we assume that the learning approach will benefit from ``knowing'' this causal structure, which is a form of domain knowledge. We used Bayesian Networks since they are a well-suited to include and model such causal domain knowledge.

\subsection{Preprocessing}

\begin{table}
\centering
  \caption{Descriptive statistics of the NaPiRE data sets \label{tab:ds-stats}}
  \begin{tabular}{lrr}
    \toprule
    Parameter & 2014  & 2018\\
    \midrule
    Participants              & 228 & 488 \\
    \ldots from Africa        & 0   & 3 \\
    \ldots from Asia          & 0   & 24 \\
    \ldots from Europe        & 126 & 208 \\
    \ldots from North America & 28  & 40 \\
    \ldots from South America & 74  & 185 \\
    \ldots from elsewhere     & 0   & 28 \\
    \midrule
    Group size           &  company      &  team \\
    \ldots \(\;x \leq 50\)      & 69     & 443\\
    \ldots \(\;51 \leq x \leq 250\)      & 33     & 39 \\
    \ldots \(\;251 \leq x\)     & 114    & 4 \\
    \ldots unknown              & 2      & 2 \\
    \midrule
    Development method          &        & \\
    \ldots agile                & 92     & 194 \\
    \ldots hybrid               & 58     & 161 \\
    \ldots plan-driven          & 46     & 124 \\
    \ldots unknown              & 22     & 0 \\
    \bottomrule
\end{tabular}
\end{table}

\begin{table}
\centering
  \caption{Available problems, causes, and effects per data set \label{tab:ds-main}}
  \begin{tabular}{lllrr}
    \toprule
     & Source & \(V\) & 2014  & 2018\\
    \midrule
    problems & predefined          & \texttt{P}  & 21 & 20 \\
    causes   & coded      & \texttt{C}  & 92  & 120 \\
    cause categories\textsuperscript{1} & predefined  & \texttt{CC} & 5 & n/a \\
    effects & coded       & \texttt{E}  & 49 & 55 \\
    effect categories\textsuperscript{2} & predefined & \texttt{EC} & 5 & n/a \\
    \bottomrule
\end{tabular}\\[1ex]

\begin{minipage}{0.75\columnwidth}
  \footnotesize
  \textsuperscript{1}Input, Method, Organization, People, Tools\\
  \textsuperscript{2}Implementation, Organization, Product, Customer, Validation
\end{minipage}
\end{table}

The 2014 NaPiRE data was obtained from Figshare, as proposed by the project~\cite{mendezNaPiREDataSet2018}. On request, the NaPiRE team provided the most recent data from the 2018 edition of the survey. A short summary of both data sets in terms of descriptive statistics is available in Table~\ref{tab:ds-stats}; for more details, please refer to M{\'e}ndez Fern{\'a}ndez~et~al.~\cite{mendezfernandezNamingPainRequirements2017}. Both consist of a set of context factor variables for each subject, which was generated from closed questions, and five answers to the problems-causes-effects question: the participants were asked to think of a recent project and to select five problems experienced in the project from a closed list. Afterwards, they were asked to assign a rank \(r \in \{1 \ldots 5\}\), a cause, and an effect to each of these problems. The latter two were coded manually, mostly in accordance with the principles established by Grounded Theory~\cite{mendezfernandezNamingPainRequirements2017}. The 2014 data set provided coarse categories into which causes and effects were grouped by the authors. Merging both data sets would have been a natural step to increase the statistical foundation of our approach, but was unfortunately prevented by incompatibilities due to survey improvements and the distinct manual coding processes.

We assigned each of the available variables to a set \(V\) (the \textit{variable type}, as depicted in Tables~\ref{tab:ds-main} and~\ref{tab:ds-context}) and then transformed it into one or more binary variables \(v_i \in V\) according to the following variable type specific rules.

\begin{itemize}
 \item Problem, cause, and effect questions: Add one binary variable per possible answer. The variable is true if the subject selected this answer in the survey, false otherwise.
 \item Cause and effects categories: Add one binary variable per category. True if the cause selected by the subject belongs to this category, false otherwise.
 \item For each context factor, we added a variable type \(V_c\), transforming it according to the data type mentioned in Table~\ref{tab:ds-context}:
 \begin{itemize}
  \item Binary: Add one binary indicator variable.
  \item Categorical: Add one binary indicator variable per value.
  \item Ordinal: Add one binary indicator variable per value.
  \item Continuous: Discretize into a set of equiprobable intervals and add one binary indicator variable per interval.
 \end{itemize}
 \end{itemize}
 
 This resulted in eight variable types for both, the 2014 and the 2018 data set with a total of 196 (2014) or 216 binary variables (2018). Given that only 28 (2014)\footnote{2014: 5 cause, 5 problem, 5 effect, 3 context factor, 10 category nodes.} or 20 (2018)\footnote{2018: 5 cause, 5 problem, 5 effect, 5 context factor, 0 category nodes.} of the \(v_i\) can actually be true for each participant, the input data matrix is relatively sparse.

\begin{table}
\centering
 \caption{Available context factors per data set \label{tab:ds-context}}
 \begin{tabular}{lllr}
  \toprule
  2014 & \(V\) & type & indicators \\
  \midrule
  company size & \texttt{CS} & categorical & 8\\
  development method\textsuperscript{1} & \texttt{CDM} & categorical & 5\\
  distributed projects & \texttt{CD} & binary & 1\\ 
  \midrule
  2018 & & type \\
  \midrule
  team size & \texttt{CS} & continuous & 6 \\
  development method\textsuperscript{2} & \texttt{CDM} & ordinal & 5\\
  distributed project & \texttt{CD} & binary & 1\\
  quality of customer relation & \texttt{CR} & ordinal & 5 \\
  system type\textsuperscript{3} & \texttt{CT} & categorical & 3 \\
  \bottomrule
\end{tabular}\\[1ex]

\begin{minipage}{0.75\columnwidth}
  \footnotesize
  \textsuperscript{1}Waterfall, V-Model XT, Scrum, XP, RUP \\
  \textsuperscript{2}Agile, rather agile, hybrid, rather plan-driven, plan-driven \\
  \textsuperscript{3}Embedded system, business information system, hybrid
\end{minipage}
\end{table}

\subsection{Network Construction}

In this work, the graph representing a Bayesian Network is defined by its architecture \(\mathcal{A}\), which is a set of tuples of variable types. Each \( (V_i, V_j) \in  \mathcal{A}\) indicates that (1) all binary variables \(v \in V_i \cup V_j\) are contained in the graph as a node and (2) that an edge is added from each \(v_i \in V_i\) to each \(v_j \in V_j\). For example, an architecture \(\mathcal{A} = \{(\texttt{C}, \texttt{P})\}\) specifies that all cause nodes are connected to all problem nodes, edges pointing to the problem nodes. This would result in \(92 \cdot 21 = 1932\) (2014) or \(120 \cdot 20 = 2400\) (2018) edges. More nodes and edges to other variable types can be specified by appending more tuples to \(\mathcal{A}\) as long as the restrictions the inference algorithm places on the graph are respected.

These considerations show that even for simple architectures, the resulting graph quickly suffers from the curse of dimensionality, which is aggravated by the manually coded cause/effect statements in the survey leading to a high number of nodes. Learning of and inference on such models would require prohibitively large amounts of memory and CPU time, so we introduced two simple filter mechanisms to reduce complexity:

\begin{itemize}
\item \textbf{Minimum Variable Occurrence Filter:} The number of true values for each variable in the data is counted. If this number is less than \(f(V)\) (i.e., less than \(f\) subjects reported this fact), its node is excluded from the graph. With this filter, variables with very little support in the dataset can be excluded.
\item \textbf{Minimum Relation Occurrence Filter:} The number of times the two variables connected by an edge are both true is counted. If this number is less than \(g(V_1, V_2)\) (i.e., the dependence was reported by less than \(g\) subjects), the edge is excluded from the graph. With this filter, relations with very little support in the dataset can be excluded.
\end{itemize}

More fine-grained control was achieved by not summing the number of occurrences of nodes or edges, but of the inverse rank \(r_{\text{inv}} = 5 - r \) of a given cause-problem-effect triple. These heuristics worked well in our case, although it might be an option to explore more sophisticated approaches like the K2 structure learning algorithm~\cite{koskiBayesianNetworksIntroduction2009}.

In the above notation, a use case is equivalent to a set of binary variables \(V_o\) which constitutes the output of a model. The choice of \(\mathcal{A}\) is independent of the use case, as long as all variables in \(V_o\) are included in \(\mathcal{A}\). Any model can thus be applied to any use case; early experimentation has shown, however, that \(\mathcal{A}\) has a major influence on the quality of the predictions. We formally define our use cases as follows\label{sec:definition-use-cases}:

\begin{center}
Diagnostic reasoning \\[-1.5em]
\begin{align*}
 V_{D} = \{ v \in \texttt{C}\;|\;v \text{ is contained in }\mathcal{A} \}
\end{align*}

Predictive reasoning\\[-1.5em]
\begin{align*}
 V_{P} = \{ v \in \texttt{P}\;|\;v \text{ is contained in }\mathcal{A} \}
\end{align*}
\end{center}

In other words, diagnostic reasoning is the prediction of causes ($\texttt{C}$) and predictive reasoning is the prediction of problems ($\texttt{P}$).

An example, which will later be called the \textit{Survey architecture}, is inspired by the causality assumptions of the NaPiRE survey:

\begin{align*}
 \mathcal{A} = \{ (\texttt{C}, \texttt{P}), (\texttt{P}, \texttt{E}) \}
\end{align*}

It is depicted graphically in Figure~\ref{fig:example-graph} with unrealistically high filter values to allow for a readable representation including individual binary variables. The following visual representations will be limited to variable types only.

\subsection{Implementation and Reproducibility}

In order to circumvent the performance problems verbally reported by other authors~\cite{kalinowskiSupportingDefectCausal2017}, we implemented this method with Julia~\cite{bezansonJuliaFreshApproach2017}, relying on the (formidable) BayesNets.jl library~\cite{wheelerSislBayesNetsJl2019} for inference and on Graphviz for visualization~\cite{gansnerOpenGraphVisualization2000}. The algorithms are wrapped in a simple web service to allow for easy integration into RE support tools and surveys. We are committed to ensuring the reproducibility of the following results, so our code and the data are freely accessible.
We cordially invite other researchers to verify and extend this work, available at our GitHub repository\footnote{\url{https://github.com/NaPiRE/project_riskdrivenRE}}
or as a static reproduction package\footnote{\url{https://doi.org/10.6084/m9.figshare.9692963}}.

\section{Evaluation}

We validated our approach in two steps. First, to evaluate whether we produce predictions that are consistent with the data, we performed a 10-fold cross-validation (leaving out 30 samples in each iteration, Section~\ref{sec:internal-validity}). Second, we selected the architecture yielding the best performance and conducted a case study to test its applicability to real-world scenarios (Section~\ref{sec:external-validity}).

\subsection{Internal Validity}
\label{sec:internal-validity}

\begin{table*}
\centering
\caption{Results of the internal validation \label{tab:internal-eva}}
\begin{tabular}{cccccrrrrrrrr}
 \toprule
 use case & \multicolumn{2}{l}{architecture}    & dataset & \(|V_o|\)\textsuperscript{1} & \(\overline{acc}\)  & \(\overline{rec}\)  & \(\overline{pre}\)  & \(rre(5)\) & \(rpr(5)\) \\
 \midrule
 \(V_D\) & \(\mathcal{A}_0\) & Baseline                    & 2018    & 29       &  0.89    & 0.07     & 0.04     &  0.32   &  0.16   \\
 \(V_D\) & \(\mathcal{A}_1\) & Kalinowski            & \textbf{2014} & 28       &  0.83    & 0.10     & 0.02     &  0.30   &  0.10   \\
 \(V_D\) & \(\mathcal{A}_2\) & Inverse Kalinowski    & \textbf{2014} & 28       &  0.92    & 0.12     & 0.37     &  0.48   &  0.15   \\
 \(V_D\) & \(\mathcal{A}_3\) & Survey                      & 2018    & 25       &  0.89    & 0.13     & 0.23     &  0.47   &  0.21   \\
 \(V_D\) & \(\mathcal{A}_4\) & Inverse Survey              & 2018    & 25       &  0.89    & 0.14     & 0.32     &  0.53   &  0.21   \\
 \(V_D\) & \(\mathcal{A}_5\) & Simple                      & 2018    & 26       &  0.89    & 0.18     & 0.35     &  0.57   &  0.26   \\
 \(V_D\) & \(\mathcal{A}_6\) & Simple with context         & 2018    & 15       &  0.91  & 0.25 & 0.81 &  \textbf{0.88}    &  0.26   \\
 \(V_D\) & \(\mathcal{A}_7\) & Inverse Simple              & 2018    & 30       &  0.90    & 0.33     & 0.73     &  0.66   &  0.30   \\
 \(V_D\) & \(\mathcal{A}_8\) & Inverse Simple with context & 2018    & 24       & \textbf{0.93}    & \textbf{0.54}  & \textbf{0.83}  
                                                                                                        &  0.81   &  \textbf{0.38}   \\
 \midrule 
 \(V_P\) & \(\mathcal{A}_0\) & Baseline                    & 2018    & 20       &  0.71    & 0.26     & 0.20     &  0.44   &  0.39   \\
 \(V_P\) & \(\mathcal{A}_1\) & Kalinowski            & \textbf{2014} & 20       &  0.74    & 0.28     & 0.59     &  0.47   &  0.39   \\
 \(V_P\) & \(\mathcal{A}_2\) & Inverse Kalinowski    & \textbf{2014} & 20       &  0.73    & 0.30     & 0.62     &  0.50   &  0.43   \\
 \(V_P\) & \(\mathcal{A}_3\) & Survey                      & 2018    & 20       &  0.8     & 0.31     & 0.71     &  0.59   &  0.51   \\
 \(V_P\) & \(\mathcal{A}_4\) & Inverse Survey              & 2018    & 20       &  0.81    & 0.49     & 0.75     &  0.63   &  0.55   \\
 \(V_P\) & \(\mathcal{A}_5\) & Simple                      & 2018    & 19       &  0.84    & 0.57     & 0.82     &  0.70   &  0.60   \\
 \(V_P\) & \(\mathcal{A}_6\) & Simple with context\textsuperscript{2}         
                                                           & 2018    & 20       &  \textbf{0.89}    & \textbf{0.73}     
                                                                            & \textbf{0.84}     &  \textbf{0.73}   &  \textbf{0.71}  \\
 \(V_P\) & \(\mathcal{A}_7\) & Inverse Simple              & 2018    & 19       &  0.77    & 0.41     & 0.69     &  0.61   &  0.53   \\
 \(V_P\) & \(\mathcal{A}_8\) & Inverse Simple with context & 2018    & 19       &  0.82    & 0.47     & 0.80     &  0.66   &  0.63   \\
 \bottomrule
\end{tabular}\\[1ex]

%
%

\begin{minipage}{0.7\textwidth}
  \footnotesize
  \textsuperscript{1}We kept the number of output variables close to 30 (\(V_D\)) and 20 (\(V_P\)) to produce comparable metrics. Divergences are caused by the limited amount of available memory.\\
  \textsuperscript{2}To reduce training time for this particular architecture, we had to limit the number of parents per node to 15.
\end{minipage}
\end{table*}

\subsubsection{Architectures} We defined the following eight architectures to be cross-validated. Four of them are inspired by the available literature.

\begin{description}
 \item[$\mathcal{A}_1$:] Kalinowski architecture~\cite{kalinowskiSupportingDefectCausal2017}\\*[0.1cm]
 \hspace*{-1cm}\centerline{\includegraphics[height=.7cm]{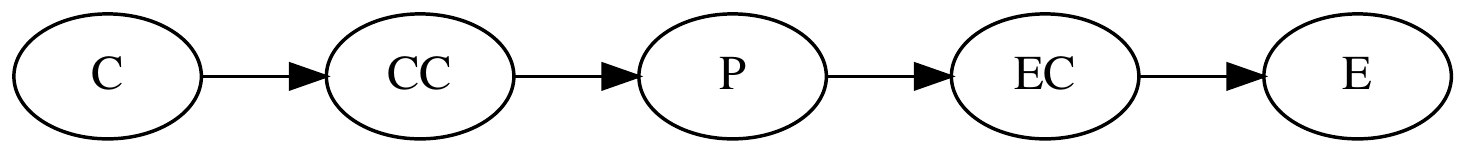}}
 \item[$\mathcal{A}_2$:] Inverse Kalinowski architecture~\cite{kalinowskiSupportingDefectCausal2017}\\*[0.1cm]
 \hspace*{-1cm}\centerline{\includegraphics[height=.7cm]{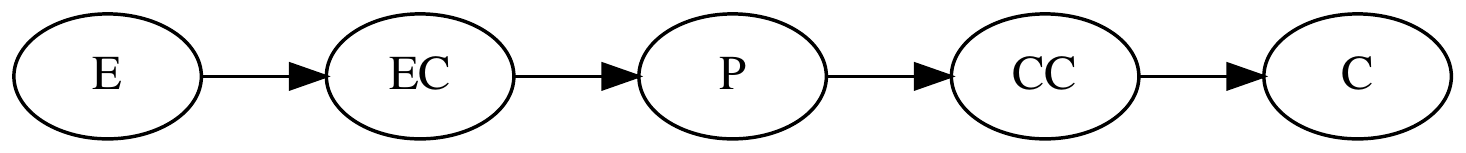}}
 \item[$\mathcal{A}_3$:] Survey architecture~\cite{mendezfernandezSupportingRequirementsEngineeringResearch2018}\\*[0.1cm]
 \hspace*{-1cm}\centerline{\includegraphics[height=.7cm]{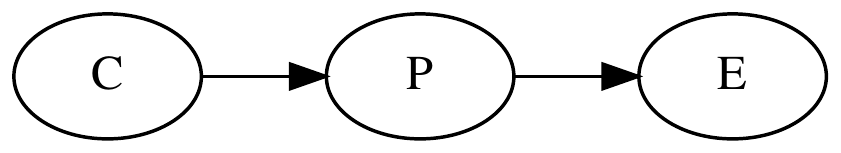}}
 \item[$\mathcal{A}_4$:] Inverse survey architecture~\cite{mendezfernandezSupportingRequirementsEngineeringResearch2018}\\*[0.1cm]
 \hspace*{-1cm}\centerline{\includegraphics[height=.7cm]{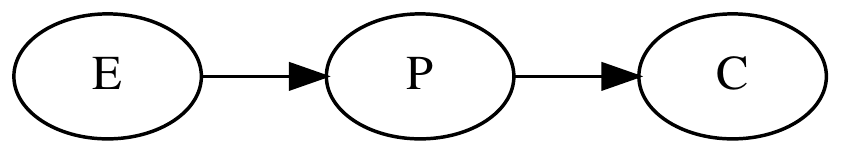}}
\end{description}

\noindent \begin{minipage}{\columnwidth}
\begin{description}[resume]
 \item[$\mathcal{A}_5$:] Simple architecture (without context factors)
 \item[$\mathcal{A}_6$:] Simple architecture (with context factors)\\*[0.1cm]
 \hspace*{-1cm}\centerline{\includegraphics[height=1.84cm]{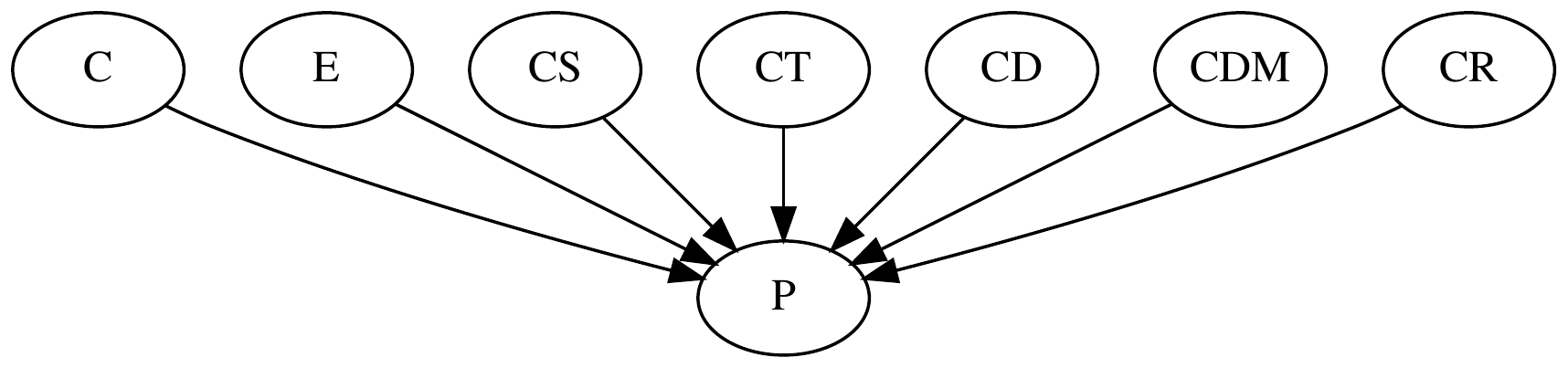}}
\end{description}
\end{minipage}

\noindent \begin{minipage}{\columnwidth}
\begin{description}[resume]
 \item[$\mathcal{A}_7$:] Inverse simple architecture (without context factors)
 \item[$\mathcal{A}_8$:] Inverse simple architecture (with context factors)\\*[0.1cm]
 \hspace*{-1cm}\centerline{\includegraphics[height=1.84cm]{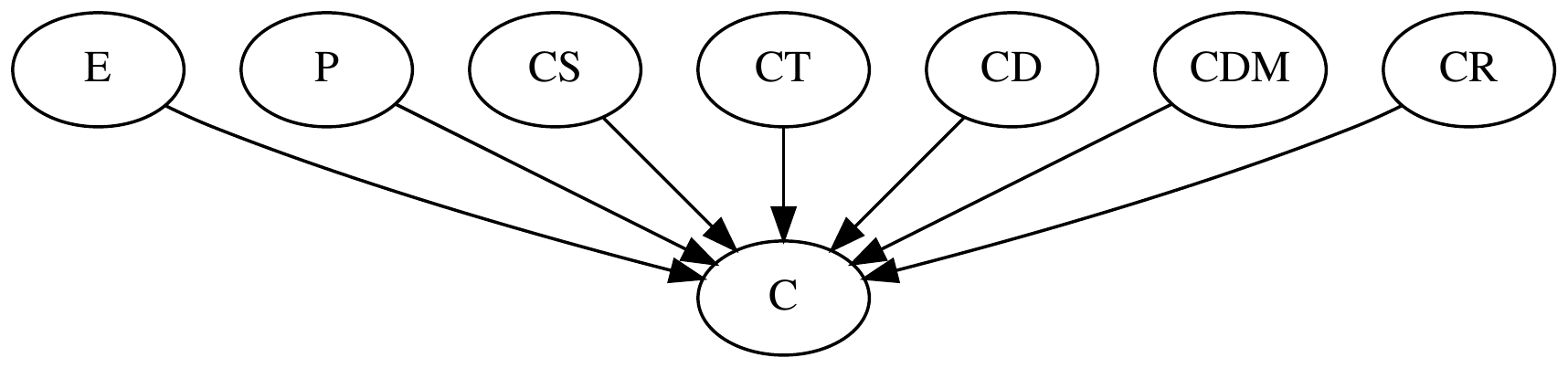}}
\end{description}
\end{minipage}\\[1ex]

To estimate whether our method achieves better performance than simple guessing does, we also defined a baseline algorithm \(\mathcal{A}_0\), predicting for each \(v_o \in V_o\) its relative frequency in the input data set.

\(\mathcal{A}_1\) and \(\mathcal{A}_2\) were evaluated on the 2014 data set since it is the only one providing the necessary cause/effect categories. The remaining architectures were evaluated on the 2018 data set due to the higher number of participants and the better availability of context factors. Filter parameters were determined empirically by repeatedly running the validation on standard hardware, incrementing the filter values until results were obtained within 48 hours and without provoking out-of-memory errors. This results in a maximum duration of \(48 \text{h} / 300 \approx 10\) minutes per inference, which we deemed the maximum acceptable inference duration for the case study.

\subsubsection{Metrics} Each tuple of architecture \(\mathcal{A}\) and use case \(V_o\) was evaluated with the following metrics. Let \(s\) be the number of samples in the validation set, \(i \in \{1, \ldots, s\}\) the sample index, and \(t = \{ 0.1, 0.2, \ldots, 0.9 \}\) a set of probability thresholds. For convenience, \(E_i(v_o)\) is the variable's actual value for the sample \(i\), \({T_i(v_o, t) = (P(V_o = v_o| E = e_i ) > t)}\) is a boolean indicator function based on the evidence \(e_i\) given by sample \(i\), and boolean values are equal to 1 or 0 in summation if they are true or false, respectively. \(V_o^k = \{ v_1, \ldots, v_k\}\) is the set of the \(k\) output variables with the highest predicted probabilities.

\begin{itemize}
 \item \textbf{binary accuracy}: the number of correct predictions to the number of all predictions.
 \begin{align*}
    acc(t) = \frac{1}{s \cdot |V_o|} \sum_{i = 1}^s \sum_{v_o \in V_o} \bigg( T_i(v_o, t) = E_i(v_o) \bigg)
\end{align*}
 \item \textbf{precision}: the number of correct predictions of true to the number of all predictions of true. 
 \begin{align*}
    pre(t) = \frac{1}{s} \sum_{i = 1}^s \left(  \sum_{v_o \in V_o | T_i(v_o, t)} E_i(v_o) \right) \left( \sum_{v_o \in V_o} T_i(v_o, t) \right)^{-1}
\end{align*}
\item \textbf{recall}: the number of correct predictions of true to the number of all actually true variables.
 \begin{align*}
    rec(t) = \frac{1}{s}  \sum_{i = 1}^s \left( \sum_{v_o \in V_o | E_i(v_o)} T_i(v_o, t) \right) \left( \sum_{v_o \in V_o} E_i(v_o) \right)^{-1}
\end{align*}
\item \textbf{ranking precision}: the number of actually true variables in the ranking to the ranking length.
  \begin{align*}
   rre(k) = \frac{1}{s} \sum_{i = 1}^s \left( \sum_{v_o \in V_o^k} E_i(v_o) \right) k^{-1}
  \end{align*}
\item \textbf{ranking recall}: the number of actually true variables in the ranking to the number of all actually true variables. 
  \begin{align*}
   rpr(k) = \frac{1}{s} \sum_{i = 1}^s \left( \sum_{v_o \in V_o^k} E_i(v_o) \right) \left( \sum_{v_o \in V_o} E_i(v_o) \right)^{-1}
  \end{align*}
\end{itemize}

As mentioned before, the dataset is relatively sparse, meaning that our baseline algorithm will correctly predict the absence of most output variables due to their low overall probability, resulting in high binary accuracy for this trivial method (known as the \emph{accuracy paradox}). Thus, this metric is less of an indicator of quality but rather points out if there are fundamental misconceptions in our approach.

Recall should be relatively high for low thresholds \(t\), but decrease as fewer variables are considered to be true. Precision should show the inverse behavior, i.e., starting low and increasing as the higher threshold filters out more false positives.

The ranking performance measures produce a good estimate of how accurate a result list similar to the one presented in Figure~\ref{fig:ui} is. There is a variety of other metrics measuring the quality of a ranking (most notably MAP, DCG, and NDCG~\cite{liuLearningRankInformation2009}), but for the external validation with non-experts, a straightforward and easy-to-interpret definition appeared more relevant to us. Recall should increase with the length of the ranking \(k\), while the precision value should decrease, which is in accordance with the canonical definitions. A notable difference is that for \(k < 5\), 100\% recall is unlikely to be achieved since the vast majority of survey participants responded with the expected 5 problem/recall/effect tuples. Because of this effect, we report \(rre(5)\), and \(rpr(5)\) instead of the respective averages.

\subsubsection{Results}

The results of our internal evaluation are presented in Table~\ref{tab:internal-eva}. All architectures except \(\mathcal{A}_1, \mathcal{A}_2\) perform clearly better than the baseline algorithm. This comparison is not entirely valid since both were evaluated on a different data set. However, our early experiments with the other architectures on the 2014 data set, which we do not present here for the sake of brevity, have hinted at a similar disparity in performance. Consequently, the effectiveness of a manual cause/effect categorization is dubious at best and the benefits should be weighed carefully against the effort required during the manual coding process.

Overall, simple models, which resemble Naive Bayesian classifiers and do not use the manifest causality assumptions implied by survey design, perform much better than models with a complex cause-effect chain. Their effectiveness can be improved further by making the \(V_o\) depend on relevant context factors, resulting in our best options \(\mathcal{A}_8\) for \(V_D\) and \(\mathcal{A}_6\) for \(V_P\), achieving good average recall/precision tuples of 0.54, 0.83 and 0.73, 0.84, respectively. The accuracy of the produced rankings appears reasonable as well, although the ranking precision of the best diagnostic model is still disappointingly low (0.38).

\begin{figure*}
  \centering
  \includegraphics[width=0.8\linewidth]{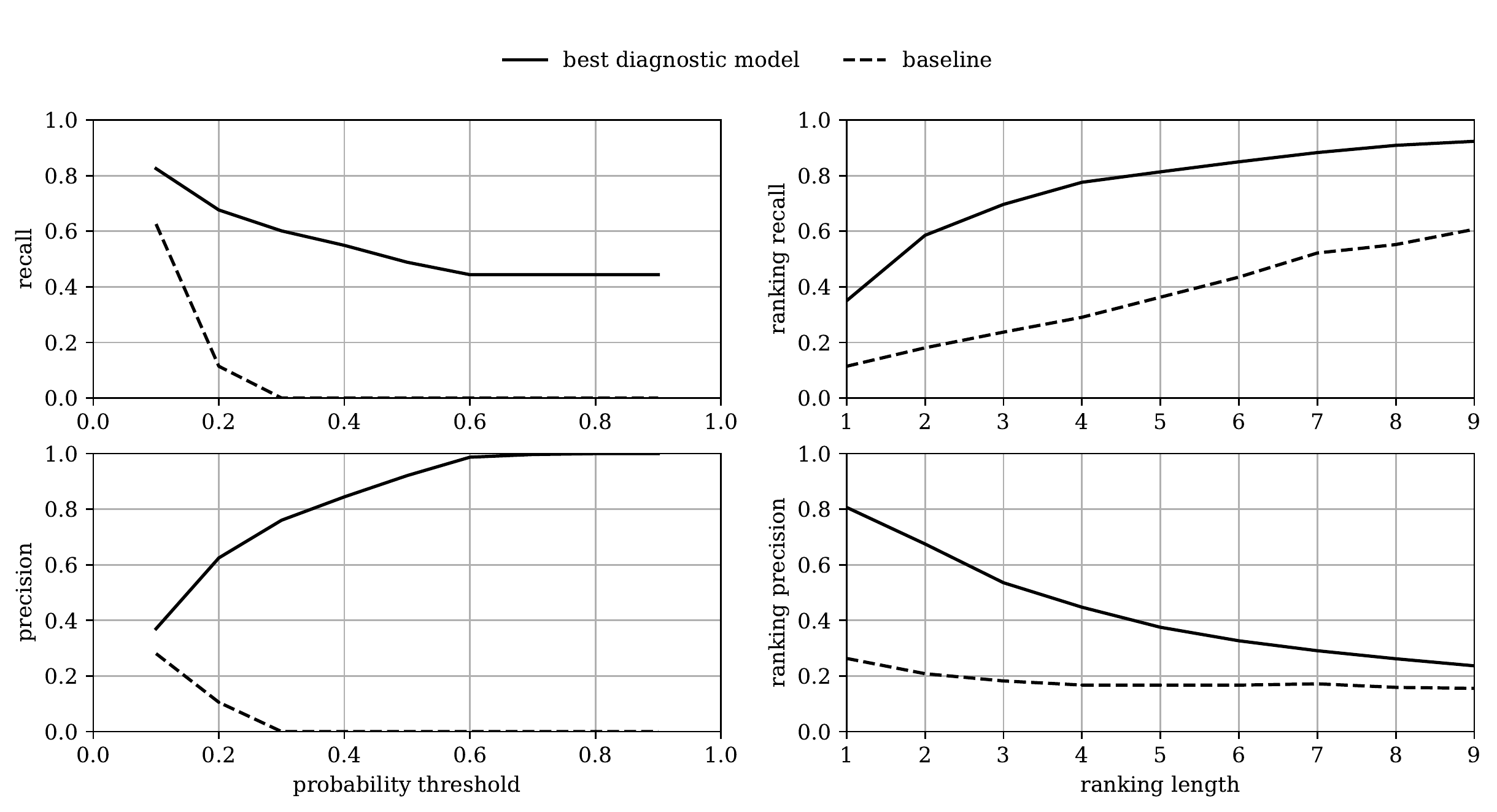}
  \caption{Metrics of the best diagnostic model \(\mathcal{A}_8\) and the baseline model.\label{fig:internal-diagnostic}}
  \label{fig:diagnostic-metrics}
\end{figure*}

\begin{figure*}
  \centering
  \includegraphics[width=0.8\linewidth]{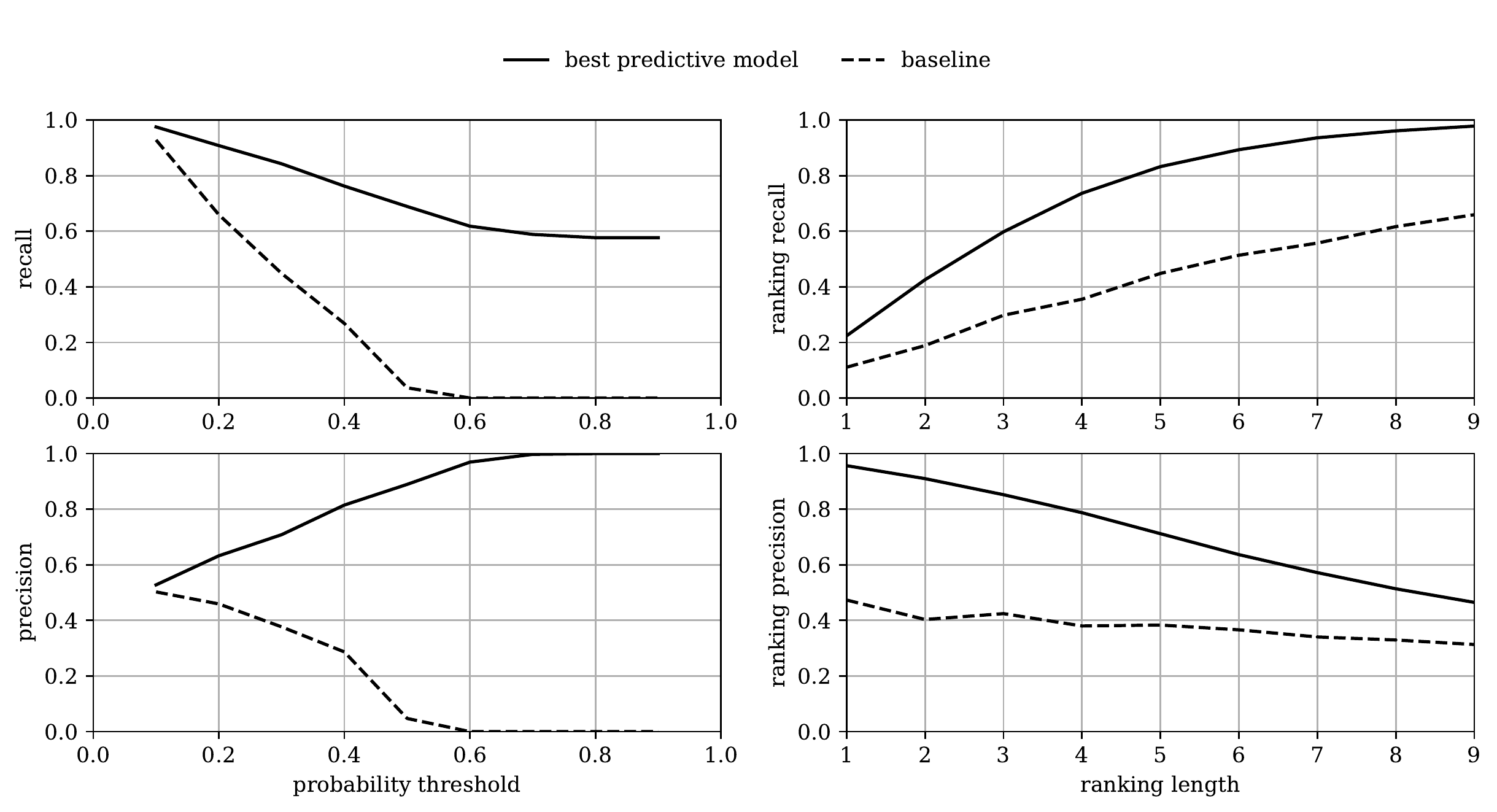}
  \caption{Metrics of the best predictive model \(\mathcal{A}_6\) and the baseline model.\label{fig:internal-predictive}}
  \label{fig:predictive-metrics}
\end{figure*}

Depending on a practitioner's needs, these metrics can be tuned by trading recall for precision and vice-versa. As visible in Figures~\ref{fig:diagnostic-metrics} and~\ref{fig:predictive-metrics}, a wide range of values is achievable. Choosing the correct trade-off for a given application has been shown to be difficult~\cite{winklerOptimizingRecallAutomatic2019}, and further investigation with a systematic study would be necessary before suggesting a specific point on these curves.

Interestingly, architectures whose edges point towards \(V_o\) perform better in many cases than architectures whose edges point away from \(V_o\): For \(V_D\), an inverse model consistently achieves better metrics than its respective non-inverted model. This is not entirely true for \(V_P\), where the Survey and Kalinowski architectures show diverging behavior. We cannot provide a sound reason for this behavior and whether it transfers to other applications remains to be seen.

Furthermore, the baseline algorithm exhibits interesting behavior: unlike expected, recall and precision fall in unison, meaning that no reasonable prediction can be achieved by simply naming the generally most probable problems or causes. Taking into account the specific circumstances of a project is thus of paramount importance for risk management.

\subsubsection{Threats to Validity}

The above results rely on a series of assumptions that need to be challenged. Our selection of architectures is by its very nature limited and does not cover all possibilities to connect the variable types. Even our definition of an architecture limits the number of Bayesian Networks which were evaluated because the basic definition does not force variables of the same type to be connected homogeneously. Furthermore, the choice of filter values by experience to reduce the computational load is, to a certain degree, arbitrary. Better or worse results could possibly be achieved by simply using different values.

Due to combinatorial effects, however, it is unlikely that an extensive evaluation will ever be possible and such computational trade-offs are a necessity. By starting with very simplistic architectures (\(\mathcal{A}_5\)--\(\mathcal{A}_8\)) whose filters we only tuned as far as necessary to be able to yield actual results, we established at least a lower bound on what should be possible to achieve on the given data sets. Adding the architectures known in the literature (\(\mathcal{A}_1\)--\(\mathcal{A}_4\)) helps to validate previously unchallenged assumptions.

The last threat emerges from the data set itself. There are no guarantees that the predefined problems and manually coded effects and causes are an internally coherent representation. Indeed, a simple random inspection of these items yielded a number of overlapping causes such as \textit{Lack of project management}, \textit{Poor project management}, \textit{Missing project management}; clearly, if one of them is present, the other causes should be present as well. In addition, there is not always a clear causality: \textit{Difficulties in project management}, which is basically synonymous to the above causes, is listed as an effect. There are other similar ambiguities in the data set whose effect is difficult to assess. 

As for any data-driven method, the quality of the dataset is extremely important (see~\cite{Vogelsang19}). Although the NaPiRE dataset is the largest of its kind that we are aware of, there are still a number of underrepresented cases. It seems that the dataset is skewed towards participants from Europe and South America, and groups with small sizes. Therefore, the performance of our approach might not be well for the minority cases from the dataset.

A minor point is that the discretization of certain context factors naturally resulted in a reduction of the precision of our input data. The break points were chosen to be equiprobable, but there is no strong evidence that these break points are an inherently good choice.

\subsection{External Validity}

\label{sec:external-validity}
\subsubsection{Research Questions}
To evaluate the external validity, i.e., to assess the validity and perceived usefulness of our approach when applied in a realistic context, we are interested in the following research questions:

\begin{itemize}
  \item {\bfseries RQ1:} How do the outputs of our tool compare with the assessment of an RE expert for a specific project?
  \item {\bfseries RQ2:} How do RE experts assess the usefulness of the tool and the style of presentation?
  \item {\bfseries RQ3:} In which contexts and for which ends would RE experts apply the tool?
\end{itemize}

\begin{figure*}
  \centering
  \includegraphics[height=6cm]{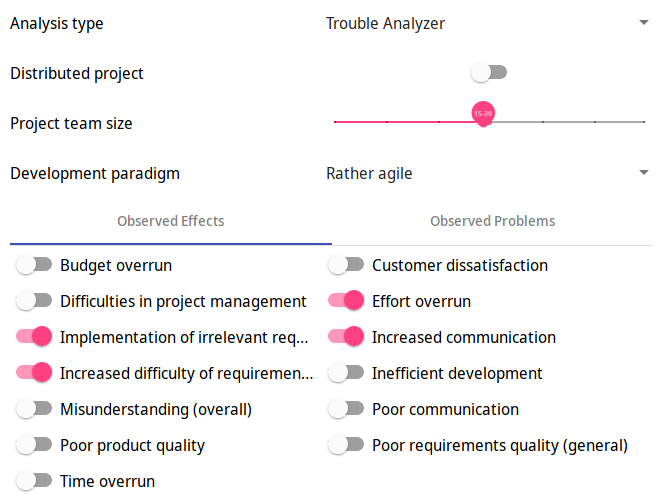}
  \includegraphics[height=6cm]{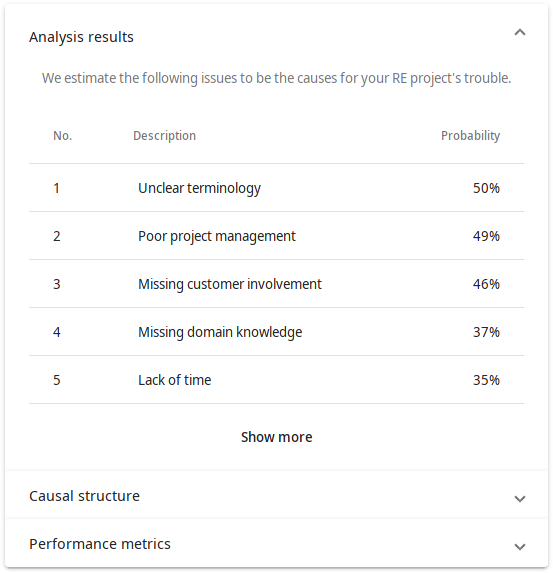}
  \caption{Data input form and output presentation (cause analysis use case)}
  \label{fig:ui}
\end{figure*}

\subsubsection{Study Design}

To answer the research questions, we conducted a case study in the context of a German software consulting company. 
We contacted an RE expert of the company and asked him whether he is interested in giving feedback on a tool for RE risk prediction. 
After he agreed, we asked him to think of a current or past RE project and offered him support for one of the two addressed use cases. 

The data collection for the case study was conducted as a semi-structured interview. Two authors and the RE expert participated in the interview that lasted around 60 minutes. The whole interview was recorded to support a detailed analysis. The interview comprised the following three parts:

\emph{Part 1:} We asked the RE expert to describe the case project, name the major problems he encountered, and list their causes according to his opinion.

\emph{Part 2:} Only now did we introduce the RE expert to our tool (as depicted in Figure~\ref{fig:ui}), backed by the model performing best in the internal validation (Section~\ref{sec:internal-validity}).
We went through the problems and effects offered by the tool and asked the RE expert whether any of them were present in the project. 
Afterwards, the tool returned a list of the 5 most likely causes together with their predicted probabilities. As additional information, the tool offered a visualization of all performance metrics and a graphical representation of the underlying graph.
We asked the expert to assess the results of the tool in terms of precision, completeness, and level of abstraction.

\emph{Part 3:} Finally, we asked the RE expert to discuss the following questions about the approach in general and the tool's particular result presentation:
\begin{enumerate}
  \item Which information presented in the user interface of our tool does the expert consider as important, what information is missing to make informed decisions?
  \item What is more relevant to the RE expert, precision of the presented results or recall? Does the expert favor a bounded list of top-X results or does he prefer a variable-length list of results above a certain probability threshold?
  \item How does the RE expert assess the impact of such data-driven predictions on his personal decisions?
  \item What is the most relevant target group for the approach? Which target groups may not benefit from the approach?
\end{enumerate}

\subsubsection{Case Description}
Our study participant is an RE consultant with 20 years of experience in general IT projects. During his early years, he worked on domain and business process modeling before focusing on testing and quality engineering. For three years, he has been working exclusively as an RE consultant.

During the case study, he was interested in analyzing a recently finished project. His role was to coach the client company's product owners for one year. The company had recently decided to move towards more agile practices.

The project itself was concerned with the enhancement of an automation portal for a ``digital factory'' with mainly automotive products. 250~team members distributed over several locations (Germany, Eastern Europe, India) were involved, a large team compared to the size in the NaPiRE dataset. The applied development process was Scrum and our study participant assessed his relation to the customer as mostly neutral with better and worse moments.

According to him, the project lends itself to a \emph{Post-Mortem-Analysis} (diagnostic reasoning) to identify the causes of several RE problems he experienced in the course of the project. 

\begin{table*}
\centering
 \caption{Problems and causes experienced by the RE expert \label{tbl:cs-participant-answers}}
 \begin{tabular}{ll}
  \toprule
  Problem & Causes\\
  \midrule
   Bad team communication   & Top-down implementation of Scrum\\
                            & No culture of failure \\
                            & Traditional company culture \\[1ex]
   Product owners were not responsible & Agile teams setup according to system components, not features. \\
                                      & Team setup was immutable\\[1ex]
   Poor requirements quality & Poor knowledge about RE \\
                             & No QA for requirements\\[1ex]
   Very technical user stories   & Poor knowledge about agile methodology\\
                                 & Silent rejection of the agile methodology \\
  \bottomrule
  \end{tabular}
\end{table*}

\begin{table}
 \centering
 \caption{Evidence provided by the RE expert\label{tbl:cs-tool-evidence}}
 \begin{tabular}{l}
  \toprule
  Evidence                                                          \\
  \midrule
  Problems                                                          \\
  \hspace{1em} Poor communication                                   \\
  \hspace{1em} Poor product quality                                 \\
  \hspace{1em} Difficulties in project management                   \\
  \hspace{1em} Misunderstandings (overall)                          \\
  \hspace{1em} Poor requirements quality (general)                  \\
  \\
  Effects                                                           \\
  \hspace{1em} Underspecified requirements                          \\
  \hspace{1em} Weak relationship between                            \\
  \hspace{3em}  customer and project team  \\
  \hspace{1em} Communication flaws within the project team          \\
  \hspace{1em} Insufficient support by customer                     \\
  \hspace{1em} Weak access to customer needs                        \\
  \hspace{1em} Incomplete or hidden requirements                    \\
  \hspace{1em} Stakeholders with difficulties                       \\
  \hspace{3em}   in separating requirements from solutions            \\
  \hspace{1em} Unclear\slash unmeasurable non-functional requirements \\
  \bottomrule
  \end{tabular}
\end{table}

\begin{table}
  \centering
  \caption{Tool predictions and expert confirmation \label{tbl:cs-tool-prediction}}
  \begin{tabular}{rlc}
  \toprule
  Rank & Cause & Conf\\
  \midrule
  1  (53\%) & Missing domain knowledge                    & \xmark\\
  2  (52\%) & Missing customer involvement                & \cmark\\
  3  (52\%) & Lack of a well-defined RE process           & \cmark\\
  4  (52\%) & Poor project management                     & \cmark\\
  5  (52\%) & Lack of time                                & \xmark\\
  6  (51\%) & Lack of requirements management             & \cmark\\
  7  (48\%) & Lack of experience of RE team members       & \cmark\\
  8  (44\%) & Communication flaws & \\
            & \hspace{1em} between team and customer & \cmark\\
  9  (31\%) & Poor requirements elicitation techniques    & \cmark\\
  10 (29\%) & Lack of communication channels              & \xmark\\
  \bottomrule
  \end{tabular}
\end{table}

\subsubsection{Study Results}
In the following, we present the results and relate them to our research questions.

{\bfseries RQ1 -- performance:} 
Table~\ref{tbl:cs-participant-answers} lists the causes and the related problems named by the interviewee. We entered the evidence reported in Table~\ref{tbl:cs-tool-evidence} into the tool. Of the predicted top-10 causes (Table~\ref{tbl:cs-tool-prediction}), our expert confirmed 7 and rejected 3. These false positives also include the cause with the highest probability (\emph{missing domain knowledge}), which was not an issue at all in the project. Besides the false positives, the expert assessed the mentioned causes as good matches. Especially, \emph{Lack of a well-defined RE process} was a top match. He also confirmed the cause \emph{Poor project management}. However, he considered this cause to be too coarse-grained without a more detailed definition, which is not provided by the data set.

{\bfseries RQ2 -- usefulness:}
The general usage of the tool was considered straightforward and the presentation of the results as a ranking was perceived to be very appropriate. The expert judges the presentation of five items on the list as a good choice because it is long enough to allow a variety of possible causes to be presented while still being tractable in group discussions. For this reason, the precision of these five predictions is of high importance, much more than achieving high recall and covering the majority of all causes that might be present.

The presentation of probabilities along the ranking was perceived as a good means to communicate the approximate risk of following the tool's suggestions, although we should have stated more clearly that such a device cannot replace a fully-fledged RE process assessment and only provides very general hints at what might be going wrong in a project. When presenting the option to show the predicted probabilities for all causes, we observed an interesting effect: every item on this long list was interpreted as a prediction of the tool by the expert, regardless of the accompanying probability.

We concluded that, while a ranking of a given length is the preferable way to display the inference results, a hybrid presentation approach is more suitable: the list should be cut off at a given probability threshold so users are not tempted to consider items with an evidently low probability. 

{\bfseries RQ3 -- applicability:} 
The RE expert proposed three primary target groups using the tool for the following purposes.

\begin{itemize}
 \item Classic projects with a project lead: Discussion input to improve the development process,
 \item Agile teams: Discussion input to improve the development process, e.g., during the Scrum Retrospective,
 \item Teams without or with inexperienced requirements engineer: Highlighting of low-hanging fruits to develop a more sophisticated RE process.
\end{itemize}

These scenarios align with his perception that the tool's results were most likely to influence group decisions, and are less likely to be able to have a tangible effect on an individual's judgments.

\subsubsection{Threats to Validity}

The above paragraphs must be seen in the light of the chosen study design: the results of case studies are inherently difficult to generalize. Despite carefully choosing an experienced participant bringing insights from a variety of projects in Software and Requirements Engineering, we conclude from the case description that the findings are probably biased towards agile methodology in larger companies and that experiences in other contexts may differ.

Another issue of case study designs is psychological bias, whose abundance forces us to focus on a selected few. Courtesy bias is hard to exclude during an interview, so most probably the above results judge our tool more positively than justifiable. Consistency bias cannot be ruled out, either. We separated the manual analysis of the situation (\textit{part 1} of the interview) from the tool introduction (\textit{part 2}), but participants would still strive to interpret the tool's predictions in a way that would produce a consistent description of the situation. In particular, generic items like \textit{Poor project management} lend themselves to such adjustments. We estimate that groupthink is less of an issue due to the participant's relatively independent position as a consultant in the project.

In the future, these concerns should be addressed by (1)~interviewing RE experts from other backgrounds to increase our coverage of different contexts, and (2)~making the tool generally available online and combining it with a questionnaire to allow anonymous feedback without a human interviewer to alleviate psychological biases.

\section{Conclusions}

Based on the NaPiRE data set, we trained a series of Bayesian Networks to model cause-effect relationships in RE projects with different contextual characteristics. These models were firstly used to conduct a post-mortem analysis, deriving probable causes of sub-optimal RE performance, and secondly to conduct a preventive analysis, predicting probable issues a young project might encounter. The method was subject to a rigorous cross-validation procedure for both use cases before assessing its applicability to real-world scenarios with a case study.

Generally, the results are promising. For both use cases, we achieve good recall and precision values with simple network architectures neglecting the causal structure implied by the underlying data set. The same is true for the quality of probability-based rankings of predicted items produced by the networks, except for the precision of the rankings for the post-mortem analysis. 

The case study involving a user-friendly interface to these models is equally supportive. The predicted causes generally matched the causes predicted by the interviewed RE expert and the presentation as a ranking was perceived as useful, although minor improvements remain. Precision was determined to be the driving performance metric in this context; unfortunately, this is the one metric in which our models perform sub-optimally. We identified a number of applications for our tool: In both, classic and agile projects its predictions can serve as valuable discussion input to improve the RE process. Moreover, it can help inexperienced teams to focus on the most worthwhile RE process and technique enhancements.

The internal validation questions the causality assumptions behind the design of the problems/causes/effects section in the NaPiRE survey: the fact that neglecting them yields considerably better performance puts into question whether what survey participants qualify as a cause actually is a cause and whether what they qualify as an effect actually is an effect. Given that distinguishing these is a notoriously difficult task even for scientists (with a large number of theoretical approaches and a variety of pitfalls such as spurious correlation), it might be too much to ask survey participants to always identify the causal chain correctly. 

Loosening these assumptions in the survey, i.e., asking for correlations only, possibly supplying a predefined list of causes and effects instead of the manual coding effort, and using undirected graphical models with a less strict structure should help improve prediction quality. On the other hand, the fact that the causality assumption seems to play only a minor role opens the field for applying other learning approaches that do not rely on this kind of domain knowledge (e.g., neural networks~\cite{Winkler16} or k-nearest neighbors).

\bibliographystyle{IEEEtran}
\bibliography{references}

\end{document}